\documentclass[epj]{webofc}
\usepackage[varg]{txfonts}   
\usepackage{epsfig}
\usepackage{graphicx}
\usepackage{mathptmx}      
\usepackage{bm}

\def\tr{{\rm tr}}

\newcommand{\SU}{{\rm SU}}

\newcommand{\vx}{{\bm{x}}}
\newcommand{\vp}{{\bm{p}}}

\newcommand{\ignore}[1]{}

\newcommand{\cL}{{\cal L}}

\wocname{EPJ Web of Conferences}
\woctitle{INPC 2013}
\begin{document}
\title{Constituent Quarks and Gluons, Polyakov loop and the Hadron Resonance Gas Model~\thanks{Presented by E.~Meg\'{\i}as at the International Nuclear Physics Conference INPC 2013, 2-7 June 2013, Firenze, Italy.} 
\fnsep
\thanks{Supported by Plan Nacional de Altas Energ\'{\i}as (FPA2011-25948), DGI (FIS2011-24149), Junta de Andaluc\'{\i}a grant FQM-225, Spanish Consolider-Ingenio 2010 Programme CPAN (CSD2007-00042), Spanish MINECO's Centro de Excelencia Severo Ochoa Program grant SEV-2012-0234, and the Juan de la Cierva Program.}
}
%

\author{E.~Meg\'{\i}as\inst{1}\fnsep\thanks{\email{emegias@ifae.es}} \and
        E.~Ruiz Arriola\inst{2} \and
        L.L.~Salcedo\inst{2}
}

\institute{Grup de F\'{\i}sica Te\`orica and IFAE, Departament de F\'{\i}sica, Universitat Aut\`onoma de Barcelona, Bellaterra E-08193 Barcelona, Spain
\and
          Departamento de F{\'\i}sica At\'omica, Molecular y Nuclear and Instituto Carlos I de F{\'\i}sica Te\'orica y Computacional, Universidad de Granada, E-18071 Granada, Spain
          }

\abstract{Based on first principle QCD arguments, it has been argued
  in~\cite{Megias:2012kb} that the vacuum expectation value of the Polyakov
  loop can be represented in the hadron resonance gas model. We study this
  within the Polyakov-constituent quark model by implementing the quantum and
  local nature of the Polyakov loop~\cite{Megias:2004hj,Megias:2006bn}. The
  existence of exotic states in the spectrum is discussed.}
\maketitle
\section{Introduction}
\label{intro}

The Hadron Resonance Gas Model (HRGM) describes the confined phase of
the QCD equation of state as a multicomponent gas of non-interacting
massive stable and point-like particles~\cite{Hagedorn:1984hz}, which
are usually taken as the conventional hadrons listed in the review by
the Particle Data Group. The Polyakov loop, a commonly used order
parameter for the hadron--quark-gluon
crossover~\cite{Borsanyi:2010bp,Bazavov:2013yv}, admits a hadronic
representation, as shown recently in~\cite{Megias:2012kb}
\begin{equation}
\langle \tr_c \Omega_3 \rangle = \langle \tr_c {\sf P} e^{i \int_0^{1/T} \! A_0\, dx_0 } \rangle  \approx \frac{1}{2} \sum_\alpha g_{h\alpha} \, e^{-\Delta_{h\alpha} / T} \,, \qquad \Delta_{h\alpha} = M_{h\alpha} - m_h \,, \label{eq:hrgm_PL}
\end{equation}
where $M_{h\alpha}$ are the masses of hadrons with exactly one heavy quark of
mass $m_h\to\infty$, and $g_{h\alpha}$ are the degeneracies of the states.
This model has been used to extract the spectrum of hadrons from a fit of
lattice data for the renormalized Polyakov
loop~\cite{Megias:2012kb,Bazavov:2013yv}. Recent lattice
data~\cite{Borsanyi:2010bp,Bazavov:2013yv} seem to indicate that conventional
hadrons are not enough to reproduce the data, and this could signal the
possible existence of exotic multipartonic states. In this communication we
will study the realization of the HRGM within a particular constituent quark
model coupled to the Polyakov loop, and address it as a diagnostic tool for
the possible existence of non conventional hadrons.

\section{The Polyakov-constituent quark-gluon model}
\label{sec:Polyakov_cqm}

An effective approach to the physics of QCD at finite temperature is provided
by chiral quark models coupled to gluon fields in the form of a Polyakov
loop~\cite{Fukushima:2003fw,Ratti:2005jh,Schaefer:2007pw}.  While most works
remain within a mean field approximation, we stressed
in~\cite{Megias:2004hj,Megias:2006bn,Megias:2005qf} the need of
quantum and local features. The partition function motivated in these works
is given by
\begin{equation}
Z = \int \prod_{\vx} d\Omega(\vx) \, e^{-S_{\textrm{PCM}}(\Omega,T)} \,,
\qquad
S_{\textrm{PCM}} (\Omega,T) =  S_q (\Omega,T) +  S_g (\Omega,T) \,,
\label{eq:3}
\end{equation}
where the Polyakov loop matrix $\Omega(\vx)$ is kept as a quantum and local
degree of freedom. $d\Omega(\vx)$ is the invariant $\SU(N_c)$ group
integration measure ($N_c=3$) at each point $\vx$.  The action of the model
depending on the quarks is obtained from the corresponding fermion determinant
and reads
\begin{equation}
S_q (\Omega,T) = -2 N_f \int \frac{d^3 x d^3 p}{(2\pi)^3} \bigg[ \tr_c
\log \big( 1+ \lambda\xi\Omega_3 (\vx) \, e^{-E_q/T}\big) + \tr_c \log \big(
  1 + \lambda\xi^{-1}\Omega_{\bar{3}}(\vx) \,e^{-E_q/T}\big) \bigg] \,,
\label{eq:quark-action}
\end{equation}
where $E_q = \sqrt{\vp^2+M_q^2}$ is the quark energy, $M_q$ is the constituent
quark mass, and $\Omega_{3(\bar{3})}$ is the Polyakov loop in the
(anti)fundamental representation. We have introduced the parameter $\lambda$
that counts the number of constituents (quarks plus antiquarks), and $\xi$ for
the number of quarks minus antiquarks. One can always replace $\lambda, \xi
\to 1$ at the end.  After a series expansion in Eq.~(\ref{eq:quark-action}),
the quark Lagrangian reads
\begin{equation}
\cL_q(\vx) = 2 N_f T \sum_{n=1}^\infty \frac{(-\lambda)^n}{n} J_n(M_q,T)
\left[ \tr_c (\Omega_3^n(\vx)) \xi^n + \tr_c (\Omega_{\bar{3}}^{n}(\vx))
  \xi^{-n} \right] \,, \label{eq:Lqexpansion31}
\end{equation}
where we have defined the functions
\begin{equation}
J_n(M,T) := \int \frac{d^3p}{(2\pi)^3} \, e^{-n E_p/T } = \frac{M^2 T}{2 \pi^2 n} K_2(n M/T) \stackrel{T \ll M}{\simeq} \left(\frac{MT}{2\pi n} \right)^{3/2} e^{-nM/T} \,, \label{eq:Jn}
\end{equation}
$K_2(x)$ being the Bessel function. One can identify in the asymptotics of
$J_n$ the statistical Boltzmann factors appearing in the low $T$ expansion of
observables. The quark propagator behaves as $\sim e^{-M/T}$, so each Boltzmann
factor is characteristic of a single quark state~\cite{Megias:2004hj}. Mesonic
contributions are identified by terms $\sim \lambda^2 \xi^0$ and they behave
as $\sim e^{-2M/T}$, while baryonic contributions are identified by terms
$\sim \lambda^{N_c}  \xi^{n N_c} $, and they behave as $\sim e^{-N_c M/T}$.

A convenient model for the gluonic action $S_g(\Omega,T)$ is~\cite{Sasaki:2012bi}
\begin{equation}
S_g(\Omega,T) = 2 \int \frac{d^3x d^3 p}{(2\pi)^3} 
\tr_c \log \left(1 - \lambda\Omega_8  \, e^{-E_g/T}  \right) \,, \label{eq:Sg}
\end{equation}
where $E_g = \sqrt{\vp^2 + M_g^2}$, $M_g$ represents a constituent gluon mass,
$\Omega_8$~is the Polyakov loop in the adjoint rep, and now $\lambda$ counts
the number of gluons. After performing the trace in color space, one
gets~\cite{Sasaki:2012bi}
\begin{equation}
\cL_g(\vx) = 2 T \int \frac{d^3p}{(2\pi)^3} \log \left( 
\sum_{n=0}^8 \lambda^n \gamma_n e^{-n E_g/T} \right) 
\,, 
\label{eq:Lg2}
\end{equation}
where $ \gamma_0 = \gamma_8 = 1\,,\; \gamma_1 = \gamma_7 = 1 - l_3 l_{\bar{3}}
\,,\; \gamma_2 = \gamma_6 = 1 - 3 l_3 l_{\bar{3}} + l_3^3 + l_{\bar{3}}^3 \,,
\; \gamma_3 = \gamma_5 = -2 + 3 l_3 l_{\bar{3}} - (l_3 l_{\bar{3}})^2 \,, \;
\gamma_4 = 2 [ -1 + l_3 l_{\bar{3}} - l_3^3 - l_{\bar{3}}^3 + (l_3
  l_{\bar{3}})^2 ] \,, \; $ and $l_\mu = \tr_c \Omega_\mu$ is the trace of the
Polyakov loop in the representation~$\mu$.

\section{Quantum and local features of the Polyakov loop}
\label{sec:quantum_local}

At low $T$, the gluonic action is small and $\Omega$ at any
given point $\vx$ is randomly distributed over the gauge group. A
convenient model to account for correlations of two Polyakov loops at
different points is 
$ \langle \tr_c \Omega (\vx) \; \tr_c \Omega^{-1}
(\bm{y}) \rangle_{S_g} = e^{-\sigma |\vx-\bm{y}| /T}
\label{eq:corr-func}
$,
with $\sigma$ the string tension.\footnote{This formula is consistent with the
  group identity $\int d\Omega \, \tr_c \Omega \, \tr_c \Omega^{-1} =1 $.}
Different values of the spatial coordinate are suppressed due to screening of
the color charge, and this defines independent confinement domains with volume
$V_\sigma = 8\pi T^3/\sigma^3$. In view of this, instead of trying to model
the higher-point correlation functions appearing in the thermal expansion, we
adopt the following approach: we assume that the space is decomposed into
domains of size $V_\sigma$, in such a way that two Polyakov loops are fully
correlated if they lie within the same domain and are fully uncorrelated
otherwise. This implies i) that each domain can be treated separately and ii)
that the Lagrangian density is $x$-independent inside each domain. Therefore,
the contribution to the partition function of any such domain is
\begin{equation}
Z = \int d\Omega \,e^{-\frac{V_\sigma}{T}(\cL_q+\cL_g)} \,.
\label{eq:Leff}
\end{equation}
This expression tells us that $e^{-\frac{V_\sigma}{T}(\cL_q+\cL_g)}$
represents the (unnormalized) probability density of the variable $\Omega$,
within the quark-gluon constituent model.

\section{Partition function}
\label{sec:Z}

By plugging Eqs.~(\ref{eq:Lqexpansion31}) and (\ref{eq:Lg2}) into
Eq.~(\ref{eq:Leff}) and performing the integration in $\SU(N_c)$, one finds
\begin{eqnarray}
&&\hspace{-1.1cm}Z \hspace{0.2cm}=\hspace{0.2cm}
1 + \lambda^2 \frac{1}{2}  \left( 2 Q_1 \bar{Q}_1 + G_1^2 + G_2  \right)
\nonumber \\ 
&&\hspace{-0.05cm}+ \lambda^3 \frac{1}{6}  \left(  Q_1^3 + 3 Q_1 Q_2
+ 2 Q_3 + \bar{Q}_1^3 + 3 \bar{Q}_1 \bar{Q}_2 
+ 2 \bar{Q}_3+ 6 Q_1 \bar{Q}_1 G_1 + 2 G_1^3 + 4 G_3 \right) + {\cal O}(\lambda^4) \,, \label{eq:Z}
\end{eqnarray}
where we have defined $Q_n(T) = 2 N_f V_\sigma \xi^n J_n(M_q,T) \,$ and
$G_n(T) = 2 V_\sigma J_n(M_g,T) \,,$ for quarks and gluons
respectively. $\bar{Q}_n$ is numerically identical to $Q_n$ but it accounts
for $n$ antiquarks, i.e. $\bar{Q}_n \sim \xi^{-n}$. Each factor $Q_n$,
$\bar{Q}_n$ or $G_n$ counts as $n$ quarks, antiquarks or gluons,
respectively. For instance, the term $Q_1 \bar{Q}_1 G_1$ has a content
$[q\bar{q}g]$. We can present the result in a schematic way (including
terms~${\cal O}(\lambda^4)$)
\begin{eqnarray}
&&\hspace{-1.1cm}Z \hspace{0.2cm}\simeq\hspace{0.2cm} 
1 + g_f^2 [q\bar{q}] + 3 [g^2] + \frac{1}{6}g_f
(g_f + 1)(g_f + 2)\big( [q^3] + [\bar{q}^3] \big) + 2 g_f^2
       [q\bar{q}g] + 4[g^3] \nonumber \\ 
&&\hspace{-0.05cm}+ \frac{1}{2}g_f^2(g_f^2+1)[q^2\bar{q}^2] + \frac{2}{3}g_f(g_f^2
       - 1)\big( [q^3g] + [\bar{q}^3g]\big) + 7 g_f^2 [q\bar{q}g^2] +
       7[g^4] + {\cal O}(\lambda^5) \,,
\end{eqnarray}
where $g_f = 2 N_f$. The factor in front of each term corresponds to
the degeneracy. At order $\lambda^2$ there are contributions from
mesons $[q\bar{q}]$ and bound states of two gluons $[g^2]$, while at
order $\lambda^3$ there are contributions from baryons $[q^3]$,
antibaryons $[\bar{q}^3]$, and other bound states involving quarks and
gluons.\footnote{The identification of these quark and gluon states
  with hadrons and glueballs follow after quantization, as explained
  in~\cite{RuizArriola:2012wd}.} At any order, the model accommodates
multiparton states with the correct counting. Some of them, like
$[q\bar{q}g]$, are irreducible color clusters which correspond to
hybrids, while other, like $[q^2\bar{q}^2]$, can be suitably arranged
as a multihadron configuration.

\section{Polyakov loop}
\label{eq:L}

The vacuum expectation value of the Polyakov loop in the representation $\mu$
can be computed as
\begin{equation}
\langle l_\mu \rangle = \frac{1}{Z} \int d\Omega\,  l_\mu \,
e^{-\frac{V_\sigma}{T} (\cL_q + \cL_g)}
\,, \qquad 
l_\mu = \tr_c \Omega_\mu \,. 
\label{eq:vevln}
\end{equation} 
Following the same procedure as for the partition function, we get for the
fundamental representation
\begin{eqnarray}
&&
\langle l_3 \rangle  =
\lambda \bar{Q}_1  + \lambda^2 \frac{1}{2}  \left( Q_1^2 + Q_2 + 2 \bar{Q}_1 G_1\right)  + \lambda^3 \left(Q_1^2 + \bar{Q}_1 G_1\right) G_1 + {\cal O}(\lambda^4) \,, \nonumber \\
&&\simeq g_f [h\bar{q}] + \frac{1}{2}g_f(g_f + 1) [hq^2] + 2g_f [h\bar{q}g] +
2g_f^2[hq^2g] + 4g_f [h\bar{q}g^2] + {\cal O}(\lambda^4) \,.
\end{eqnarray}
We have identified the Polyakov loop itself with a heavy quark source~''$h$'',
which is screened by dynamical (anti)quarks and gluons from the medium. The
first two terms $[h\bar{q}]$ and $[hq^2]$ correspond to mesons and baryons,
respectively, with a heavy quark and one or several light (anti)quarks. This
completes the connection with the HRGM for the Polyakov
loop~\cite{Megias:2012kb}. It is noteworthy that tetraquark states
$[h\bar{q}^2q]$ that do appear in numerator and denominator of
(\ref{eq:vevln}), cancel in $\langle l_3 \rangle$. This implies that all the
states of the type $[h\bar{q}^2q]$ in the numerator (namely, the partition
function in presence of the fundamental source $h$) can be accounted for by
clusters of the type $[h\bar{q}][\bar{q}q]$, that is, light-heavy hybrid
mesons $[h\bar{q}]$ from straight screening of the source, plus an ordinary
meson $[q\bar{q}]$ occurring in the domain of the source. On the other hand,
the other contributions can not be arranged into two or more singlet
subclusters and should be regarded as genuine states screening the source.
The situation becomes more involved when more partons are included~\cite{mrs:prep}.

\section{Conclusions}
\label{sec:conclusions}

We have studied the low temperature regime of QCD by using a chiral
quark model with Polyakov loop. When the local and quantum nature of
the Polyakov loop is taken into account, we find a clear connection
with the hadron resonance gas description. In addition, the model
definitely contains some quark-gluon hybrid states that could be
exposed by saturation of the Polyakov loop expectation value sum rule
in Eq.~(\ref{eq:hrgm_PL}). The status of other exotic states of the
type tetraquark or pentaquark is much less clear. A high precision
computation of observables at low temperature in lattice can serve as
a powerful tool to disentangle this rich content of the QCD spectrum.

%
%
%

\end{document}